\documentclass[pre,preprint,showpacs]{revtex4}
\usepackage[english]{babel}
\usepackage{amscd}
\usepackage{graphicx}
\usepackage{amsmath}
\usepackage{amsfonts}
\usepackage{amssymb}
\usepackage[latin1]{inputenc}
\usepackage{color}

\newcommand{\be}{\begin{equation}}
\newcommand{\ee}{ \end{equation}}
\newcommand{\ben}{\begin{eqnarray}}
\newcommand{\een}{\end{eqnarray}}

\begin{document}

\title{The connection between Jackson and Hausdorff derivatives in the
  context of generalized statistical mechanics}

\author{Andre A. Marinho$^{1,2}$, G.M. Viswanathan$^{1,3}$, Francisco A. Brito$^{2,4}$, C.G. Bezerra$^{1}$}

\affiliation{$^{1}$ Departamento de Física, Universidade Federal do Rio
Grande do Norte, 59078-970 Natal, RN, Brazil
\\
$^{2}$ Departamento de Física, Universidade Federal de
Campina Grande, 58109-970 Campina Grande, Paraiba, Brazil
\\
$^3$National Institute of Science and Technology
                   of Complex Systems,
                   Universidade Federal do Rio Grande do Norte,
                   Natal--RN, 59078-970, Brazil
\\
$^{4}$ Departamento de Física, Universidade Federal da Paraíba,
Caixa Postal 5008, 58051-970 João Pessoa, Paraíba, Brazil}
\date{\today}

\begin{abstract}

In literature one can find many generalizations of the usual Leibniz
derivative, such as Jackson derivative, Tsallis derivative and
Hausdorff derivative. In this article we present a connection between
Jackson derivative and recently proposed Hausdorff derivative. On one
hand, the Hausdorff derivative has been previously associated with
non-extensivity in systems presenting fractal aspects. On the other
hand, the Jackson derivative has a solid mathematical basis because it
is the $\overline{q}$-analog of the ordinary derivative and it also arises
in quantum calculus. From a quantum deformed $\overline{q}$-algebra we
obtain the Jackson derivative and then address the problem of $N$
non-interacting quantum oscillators. We perform an expansion in the
quantum grand partition function from which we obtain a relationship
between the parameter $\overline{q}$, related to Jackson derivative,
and the parameters $\zeta$ and $q$ related to Hausdorff derivative and
Tsallis derivative, respectively.


\end{abstract}

\maketitle

\section{Introduction}
\label{int}
The observation of nature has always motivated the human curiosity
about understanding the world around us. In particular, from a
mathematical perspective, new concepts and ideas have been developed
in the last centuries to model and describe a wide variety of complex
physical systems. For example, it is not possible to associate
Euclidean geometry with many forms found in nature, such as the shape
of clouds, coastlines, etc. This difficulty in describing some
physical systems in nature is basically due to
non-standard
irregularities  and patterns that are
present. In spite of the enormous complexity, most of these forms and
trajectories present relatively simple scaling laws. The understanding
of these complex and irregular physical systems was the motivation for
the development of the concept of fractal
geometry\cite{arm,ken,man,dar}.

Fractal objects have been identified in many physical situations and play an important role in very distinct areas, such as topology, Brownian motion, fluid turbulence, surface roughness, porosity of rocks, dynamics systems, number theory, population of species, applications in medicine, quasicrystals \cite{lop,ben,dla,cgb1,cgb2}, etc. There is a close connection between fractal geometry and chaotic systems, which has allowed the possibility of seeing order and patterns where previously only the random and chaotic were observed, ranging from problems of aggregation to the behavior of chaotic dynamical systems \cite{hei}.

Fractals have peculiar properties and characteristics. As a
consequence of the irregularities found in systems presenting fractal
behavior, the tools of classical calculus may not be adequate to
address natural phenomena described by functions associated to those
systems. In particular, in the last decades several proposals for
extending the concept of calculus, with great potential of application
for the study of systems presenting fractal behavior, have
appeared. Among these proposals we can cite many generalizations of
the usual Leibniz derivative, such as the Jackson derivative, the
Tsallis derivative, fractional derivatives and the Hausdorff
derivative, which can be applied in a large set of functions.  We
highlight the Hausdorff derivative, with applications in mapping of
fractal domains into the continuous \cite{wen,ale,ale2} (for
properties of coherent states of dissipative systems and in relations
developed with generalized algebras for the study of complex systems
in statistical mechanics \cite{lyr,web2,web3,vit2}).

It is known from the literature that a possible mechanism to generate
a deformed version of the classical statistical mechanics is to
replace the usual Boltzmann-Gibbs distribution by a deformed
version. To do so, a form of deformed entropy is postulated that
implies a theory of generalized thermodynamics
\cite{tsa,kan,abe,ern,sla,aps,amf,ntc}. We can find in literature a
number of generalizations of well-known Boltzmann-Gibbs statistical
mechanics obtained from $q$-deformed algebras
\cite{jac2,bie,mac,flo,arik,lav,okt,aal3,won,oil}. In particular,
$q$-calculus in the Tsallis version of non-extensive statistical mechanics
has been applied to a wide class of different systems.


In a recent work Weberszpil {\it et al.}
showed a connection
between $q$-deformed algebra, in Tsallis version of non-extensive
statistical mechanics, with Hausdorff derivative, mapped into a
continuous medium with a fractal measure \cite{web2}. In this paper we show a
relationship between Jackson derivative, Hausdorff derivative and
Tsallis $q$-derivative.

Our interest in the Jackson derivative stems from
  the fact that it is the $\overline{q}$-analog of the ordinary derivative.
  Unlike more recent deformations of the geometric and hypergeometric
  series, the theory of $\overline{q}$-series has centuries of solid mathematical
  underpinning.  The study of $\overline{q}$-series began with some theorems of
  Gauss (e.g., for theta functions), Euler (e.g., the combinatorial
  version of the pentagonal number theorem) and Cauchy (e.g.,
  $\overline{q}$-analog of the binomial theorem). At the turn of the 19th century
  and the begininng of the 20th century, the English mathematicians
  F.~H.~Jackson (known for the JD) and L.~J.~Rogers (known for the
  Rogers-Ramanujan identities) built the foundations of this
  important research area of mathematics. It is often said that the
  person who most contributed to $\overline{q}$-series was S. Ramanujan. Given
  this long mathematical tradition, the $\overline{q}$-analogs have a far more
  solid theoretical basis in comparison to other deformations of the
  standard algebra. Our motivation here is thus to take the HD and to
  try to replace it with the better-understood JD.

This article is organized as follows. In
Sec.\ \ref{hdq} we review the connection between Hausdorff derivative
and Tsallis $q$-derivative. In Sec.\ \ref{dqa} we obtain Jackson
derivative from deformed quantum algebra. Sec.\ \ref{jdh} is devoted
to develop the relationship between Jackson and Hausdorff
derivatives. Finally in Sec.\ \ref{con} we make our final comments.

\section{Connection between Hausdorff and Tsallis $q$-derivatives}
\label{hdq}
Statistical mechanics is one of the areas of physics that deals with the so-called complex systems, which have been attracting a lot of attention in the last decades. In order to approach such complex systems, in the last years several non-conventional formalisms have been proposed, among them we can point out Tsallis, Kaniadakis and Abe \cite{tsa,kan,abe}, which are generalizations of Boltzmann-Gibbs statistics \cite{sal,patt} through modified entropies. On the other hand, formulations of deformed derivatives have shown great potential of applications in the study of complex systems and phenomena presenting fractal properties. In this sense, the concept of Hausdorff derivative (HD) of a function, with respect to a fractal measure \cite{wen}, has demonstrated possible connections with the $q$-derivatives in the framework of non-extensive statistics \cite{web2,web3,vit2}, leading to a better understanding of both formalisms. It is worth to remark that HD differs from the standard fractional derivative because it does not involve the convolution integral and it is local in nature.

Let us consider HD derivative which is defined by \cite{wen},
\be \frac{d^H}{dx^\zeta}f(x)=\lim_{x\to y}\frac{f(y)-f(x)}{y^\zeta-x^\zeta},\ee
with the local fractional differential operators being described as \cite{ale},
\be \frac{d^H}{dx^\zeta}f(x)=\left(1+\frac{x}{l_0}\right)^{1-\zeta}\frac{d}{dx}f=
\frac{l_0^{\zeta-1}}{c_1}\frac{d}{dx}f=\frac{d}{d^{\zeta}x}f.\ee
\noindent Here $l_0$ is the lower cutoff along the cartesian $x$-axis and the scaling expoent $\zeta$ characterizes the density of states along the direction of the normal to the intersection of the fractal continuum with the plane (see Ref.\ \cite{ale} for details). Following the lines of Ref.\ \cite{web2}, we perform an expansion in $x$ to the first order with exponent $(1-\zeta)$. We get,
\be \label{eq1} \frac{d^H}{dx^\zeta}f(x)=\left(1+\frac{x}{l_0}\right)^{1-\zeta}\frac{d}{dx}f(x)\approx
\left[1+\frac{(1-\zeta)}{l_0}x+\cdots\right]\frac{d}{dx}f(x).\ee

Let us now consider the $q$-deformed derivative. In the Tsallis non-extensive framework, the $q$-deformed difference operator is defined by \cite{ern},
\be x\ominus_q z = \frac{x-z}{1+(1-q)x},\qquad\qquad \mbox{with $z\neq$$\left(\frac{1}{(q-1)}\right)$}.\ee
\noindent Therefore, we may write the Tsallis $q$-derivative as
\be \label{eq2} D_{(q)}f(x)=\lim_{z\to x}\frac{f(x)-f(z)}{x\ominus_q z}=[1+(1-q)x]\frac{d}{dx}f(x).\ee

By comparing Eqs.\ (\ref{eq2}) and (\ref{eq1}), Weberszpil and collaborators \cite{web2} obtained a relationship between Tsallis $q$-derivative and HD derivative as follows,
\be 1-q = \frac{1-\zeta}{l_0}.\ee

\noindent Those authors conclude that the deformed Tsallis $q$-derivative is the first order expansion of the Hausdorff derivative and that there is a strong connection between those formalisms by means of a fractal metric.

\section{$\overline{q}$-Deformed Quantum Algebra}
\label{dqa}
It is worth to remark that the $\overline{q}$-deformed quantum algebra, considered in the present section to introduce Jackson derivative, is a generalization of the Heisenberg quantum algebra. Therefore, it is defined in a completely different context from the Tsallis $q$-algebra (which is defined in the non-extensive statistical mechanics context) presented in the previous section. In order to avoid misunderstanding, in this work we use the notation $q$ for Tsallis algebra and $\overline{q}$ for the deformed quantum algebra.

At the beginning of the 20th Century, Jackson developed a number of works which have in many aspects played an important role in the understanding and developing of deformed quantum algebra \cite{jac2}. The appearance of the deformation is made through the key ingredient - the deformation parameter $\overline{q}$, which is introduced in the commutation relations that define the Lie algebra of the system. The original Lie algebra, not deformed, is recovered when $\overline{q}\to 1$.

Let us illustrate Jackson derivative by considering $N$ non-interacting quantum oscillators in order to define a generalized $\overline{q}$-deformed dynamics in a $\overline{q}$-commutative phase space. The algebraic symmetry of the quantum oscillator is defined by the Heisenberg algebra in terms of annihilation and creation operators $c$, $c^{\dagger}$, respectively, and the number operator $N$ as follows \cite {lav,sak},
\begin{eqnarray} \label {eq3} [c,c]_\kappa = [c^{\dagger},c^{\dagger}]_\kappa =0\quad,\qquad\qquad\ cc^{\dagger} -\kappa
\overline{q}c^{\dagger}c=1,\end{eqnarray}
\begin{equation}[N,c^{\dagger}]= c^{\dagger}\quad, \qquad\qquad [N,c] = -c. \end {equation}
\noindent Here the deformation parameter $\overline{q}$ is real; the constant $\kappa = 1$ for $\overline{q}$-bosons
(with commutators) and $\kappa = -1$ for $\overline{q}$-fermions (with anticommutators). We also define the \textsl{basic number} as
\begin{equation}\label{eq4} [x_i]=c_{i}^\dagger c_{i}=\frac {\overline{q}^{x}-1} {\overline{q}-1}, \end{equation}
with
\begin{eqnarray}[x,y]_\kappa=xy-\kappa yx\quad, \qquad\qquad cc^{\dagger}=[1+\kappa N]. \end {eqnarray}
Note that for $\overline{q}\neq 1$ the \textsl{basic number} $[x]$ does not meet additivity, i.e.,
\begin{equation} [x+y] = [x] + [y]+(\overline{q}-1)[x][y]. \end{equation}
It is also easy to observe that as $\overline{q}\to 1$ the \textsl{basic number} $[x]$ is reduced
to an ordinary number $x$.

The $\overline{q}$-Fock space spanned by the orthornormalized eigenstates $|n\rangle$ is constructed according to
\begin{eqnarray} {|n\rangle} =\frac{(c^{\dagger})^{n}} {\sqrt{[n]!}}{|0\rangle}\quad,\qquad\qquad c{|0\rangle}=0,
\end {eqnarray}
where the factorial of the \textsl{basic number} $[n]$ is defined as
\begin{equation}[n]!=[n][n-1],\cdots,[1].\end{equation}
The applications of $c$, $c^{\dagger}$ and $N$ to a state $|n\rangle$ in the $\overline{q}$-Fock space are known to provide
\begin{equation} c^{\dagger}{|n\rangle} = [n+1]^{1/2} {|n+1\rangle}, \end{equation}
\begin{equation} c{|n\rangle} = [n]^{1/2} {|n-1\rangle},\end{equation}
and
\begin {equation} N {|n\rangle} = n{|n\rangle}. \end{equation}
We may perform a linear transformation from the $\overline{q}$-Fock space to the configuration space (Bargmann Holomorphic representation) \cite{flo} as follows,
\begin{eqnarray} c^{\dagger} = x,\quad\qquad\qquad  c = \partial_{x}^{(\overline{q})}.\end{eqnarray}
Here $\partial_{x}^{(\overline{q})}$ is the Jackson derivative (JD)\cite{jac2},
\begin{equation}\label{eq5}\partial_{x}^{(\overline{q})}f{(x)}=\frac {f(\overline{q}x)- f(x)}{x(\overline{q}-1)}.\end{equation}
Note that it becomes an ordinary derivative as $\overline{q}\to 1$. Therefore, JD naturally occurs in quantum deformed systems and, as we will show later, it plays an important role in the $\overline{q}$-generalization of thermodynamics relations.

Let us address, for  example, the Hamiltonian of $\overline{q}$-deformed non-interacting quantum oscillators,
\be {\cal H} = \sum_{i}{(\epsilon_i-\mu)}{N_i}=\sum_{i}{(\epsilon_i-\mu)}{c_i^{\dagger}c_i}.\ee
Here $\mu$ is the chemical potential, $\epsilon_i$ is the kinetic energy in state $i$ associated to
the operator number $N_i$. The Hamiltonian is deformed and implicitly depends on $\overline{q}$ through the basic number defined by Eq.\ (\ref{eq4}). The mean value of the $\overline{q}$-deformed occupation number $n_{i}^{(\overline{q})}$ can be calculated by
\be [n_{i}^{(\overline{q})}]\label{eq6}\equiv \langle[n_{i}^{(\overline{q})}]\rangle = \frac{tr(\exp(-\beta{\cal H}) c_{i}^\dagger c_{i})}{\Xi}.\ee
Here $\beta=(\kappa_B T)^{-1}$, $\kappa_B$ is the Boltzmann constant, $T$ is the temperature and $\Xi=tr[\exp(-\beta{\cal H})]$
is the partition function of the system. From Eqs.\ (\ref{eq3}), (\ref{eq4}) and (\ref{eq6}), using the cyclic property of the trace \cite{lav},
we get
\ben \label{eq7}n_{i}^{(q)} = \frac{1}{\ln(\overline{q})}\ln\Bigg\{\frac
{z^{-1}\exp(\beta\epsilon_i)-1}{z^{-1}\exp(\beta\epsilon_i)-\overline{q}^\kappa}\Bigg\},\een
where $z=\exp(\beta\mu)$ is the fugacity of the system. From the algebra of non-deformed quantum oscillators we know that the average occupation number is given by
\begin{equation} \label{eq71}n_{i} = \frac{1}{z^{-1}\exp(\beta\epsilon_i)-1},\end{equation}
with
\begin{eqnarray} N=\displaystyle\sum_{i}n_{i} \qquad \mbox{and in a similar way} \qquad  N^{(\overline{q})} = \displaystyle\sum_{i}n_{i}^{(\overline{q})}. \end{eqnarray}

\noindent We can obtain the total number of particles $N$ from the logarithm of the grand partition function $\Xi$, i.e.,
\begin{equation}\label{eq8}\ln{\Xi}=-\kappa\displaystyle\sum_{i}{\ln{\left[1-z\kappa\exp(-\beta\epsilon_i)\right],}}\end{equation}
so that
\begin{eqnarray}\label{eq9} N =z\frac{\partial}{\partial z}\ln{\Xi}=\displaystyle\sum_{i}\frac{z\exp(-\beta\epsilon_{i})}
{1-z\exp(-\beta\epsilon_{i})}= \displaystyle\sum_{i}\frac{1}{z^{-1}\exp(\beta\epsilon_{i})-1}=
\displaystyle\sum_{i}n_{i}.\end{eqnarray}

\noindent However, the total number of particles in the formalism of the $\overline{q}$-deformed oscillators, $N^{(\overline{q})}$,
can not be obtained by using usual thermodynamics. On the other hand, we can establish a relationship between $N^{(\overline{q})}$ and $N$ by performing an expansion considering $z\ll 1$, corresponding to high temperature or diluted gas limit, in Eqs.\ (\ref{eq7}) and (\ref{eq71}), i.e.,

\begin{eqnarray} n_{i}^{(\overline{q})}=\frac{\overline{q}-1}{\ln(\overline{q})}z \exp(-\beta\epsilon_{i})
\qquad \mbox{and} \qquad n_{i}=z\exp(-\beta\epsilon_{i}), \end{eqnarray}
which means
\begin{equation}\label{e43}n_{i}^{(\overline{q})}=\frac{\overline{q}-1}{\ln(\overline{q})}\;n_{i}\end{equation}
and
\begin{eqnarray} \displaystyle\sum_{i}n_{i}^{(\overline{q})}=\frac{\overline{q}-{(\overline{q})}^{-1}}{2\ln (\overline{q})}\displaystyle\sum_{i}n_{i}\;\;\; \Rightarrow \;\;\;
N^{(\overline{q})}=\frac{\overline{q}-1}{\ln(\overline{q})}\; N.\end{eqnarray}

\noindent Therefore the deformed version of Eq.\ (\ref{eq9}) may be written as
\begin{equation} N^{(\overline{q})}=zD_{z}^{(\overline{q})} \ln{\Xi}=\frac{\overline{q}-{(\overline{q})}^{-1}}{2\ln (\overline{q})}\, z\frac{\partial}{\partial z}\,\ln{\Xi}. \end{equation}
Here $D_{z}^{(\overline{q})}$ is the so-called deformed differential operator defined as
\ben\label {eq10} D_{z}^{(\overline{q})}=\frac{\overline{q}-1}{\ln(\overline{q})}\;
\partial_{z}^{(\overline{q})} \qquad\mbox{with limit $\overline{q}\to 1$}\qquad \;\; \Rightarrow \;\; D_{z} = \frac{\partial}{\partial z}.\een
It is quite clear from Eq.\ (\ref{eq10}) the connection between deformed derivative $D_{z}^{(\overline{q})}$ and the usual derivative defined by Leibniz $\frac{\partial}{\partial z}$.

\section{Connection between JD and HD}
\label{jdh}
In order to obtain a connection between $JD$ and $HD$ we follow an approximation similar to that made in Eq.\ (\ref{eq1}) of Sec.\ \ref{hdq}. Thus, we will carry out a second order expansion of Eq.\ (\ref{eq5}), with $f(x)=\exp[-\beta(\epsilon-\mu)]$. We get,
\be\label{eq11}\partial_{x}^{(\overline{q})}f(x)\approx\left[1+\frac{(\overline{q}+1)}{2}x+\cdots\right]\frac{d}{dx}f(x).\ee
By comparing Eqs.\ (\ref{eq1}) and (\ref{eq11}) we can infer that
\be\label{eq12} \frac{\overline{q}+1}{2}=\frac{1-\zeta}{l_0}.\ee
From Ref.\ \cite{web2} we know that
\be 1-q = \frac{1-\zeta}{l_0},\ee
so that we can relate both deformation parameters $q$ and $\overline{q}$, as follows
\be \overline{q} = 1 - 2q.\ee

\noindent Therefore, now we have the deformed parameter $\overline{q}$ connected to the fractal metric, via $\zeta$, and to the non-extensive Tsallis framework, via $q$. This result provides the recipe for  using the JD to substitute for the HD, which has the advantages previously discussed of having a solid theoretical foundation. It is a also a clear indication that the HD does not have any unique property that renders it fundamental importance, rather the JD is able to play the same role.

Let us discuss about our result. It is well known that the Boltzmann factor $e^{-\beta
    \epsilon} $ invariably arises in the treatment of the
  thermodynamic limit of isolated systems in equilibrium whose
  Hamiltonians have sufficiently short-range interactions. However, if
  the systems have finite size or are not in equilibrium or else if
  the Hamiltonian has long-range interactions, then there is no reason
  to expect that the Boltzmann factor will properly describe the
  energy distribution. In this context, alternative formulations that
  generalize the Boltzmann factor have been proposed, whereby the
  exponential function suffers a deformation.  Recall that the
  exponential function $\exp(\lambda x )$ is an eigenfunction of the
  differential operator $d/dx$ with eigenvalue $\lambda$. In
  contrast, a deformed exponential will not, in general, be an
  eigenfunction of $d/dx$. For example, the Tsallis $q$-exponential
  $\exp_q(x) $ is not an eigenfunction of $d/dx$ except when $q=1$.
  It is thus natural to expect that the ordinary derivative should be
  replaced by other derivatives, such as the Hausdorff
  derivative~\cite{web2,web3,vit2}.  But there is more than one way to
  deform the exponential function.  For example, in addition to the
  Tsallis $q$-exponential function, we can cite the
  Kaniadakis~\cite{kan} $\kappa$-exponential function
  $\exp_\kappa(x)$.  One of the oldest deformations is the $q$-analog
  $e_q(x)$ of the exponential function (note that $e_q(x) \neq
  \exp_q(x)$, the latter being the Tsallis $q$-exponential). It is in
  this context that we can use the JD, because the JD is the $\overline{q}$-analog of
  the ordinary derivative.  Our results show that the JD, with its
  deeper mathematical origins, can be just as useful as the HD.  Since
  the HD has been applied to study fractal and other systems where the
  Boltzmann-Gibbs treatment is expected to not be applicable, we expect
  that the JD will be similarly useful.


\section{Conclusions}
\label{con}

Fractal systems are relevant in many areas of science. Their
statistical description and phenomenological modeling lead to a new
understanding of complexities and irregularities of nature. In order
to address systems presenting fractal geometry, Hausdorff derivative (HD)
has been considered and applied. In a similar way, Tsallis
$q$-derivative has been considered and applied in the study of
non-extensive statistical mechanics systems. Recently, Weberszpil and
collaborators \cite{web2} obtained a relationship between
$q$-derivative and HD derivative, $1-q = \frac{1-\zeta}{l_0}$, showing
that there is a strong connection between those formalisms by means of
a fractal metric. On the other hand, Jackson derivative (JD) is a
generalization of the Heisenberg quantum algebra and it is connected
to the study of $\overline{q}$-deformed quantum systems. In the present work we
showed a relationship between $\overline{q}$-derivative and HD
derivative, $\frac{\overline{q}+1}{2}=\frac{1-\zeta}{l_0}$, and
between $\overline{q}$-derivative and $q$-derivative, $1-q =
\frac{1-\zeta}{l_0}$. In particular, the deformation parameters $q$
and $\overline{q}$ are connected by $\overline{q} = 1 - 2q$. We can
conclude that the deformed parameter $\overline{q}$ is connected to
the fractal metric, via $\zeta$, and to the non-extensive Tsallis
framework, via $q$. Our results show that the JD, with its
  deeper mathematical origins, can be just as useful as the HD.  Since
  the HD has been applied to study fractal and other systems where the
  Boltzmann-Gibbs treatment is expected to not be applicable, we expect
  that the JD will be similarly useful. We believe that our results may open new
perspectives on the understanding of complex systems.


\section*{Acknowledgments}

We would like to thank CNPq, CAPES, and PNPD/CAPES, for partial financial support. FAB acknowledges support from
CNPq (Grant no. 312104/2018-9).


\end{document}